\documentclass[12pt]{iopart}

\usepackage{iopams}  
\usepackage{cite}
\usepackage{color}
\usepackage{graphicx}
\begin{document}

\title[Pulse length effects in long wavelength driven non-sequential double ionization]{Pulse length effects in long wavelength driven non-sequential double ionization}

\author{H. Jiang$^{1}$\footnote{These authors contributed equally}, M. Mandrysz$^{2}\ddagger$, A. Sanchez$^{3}$, J. Dura$^{3}$, T. Steinle$^{3}$, J. S. Prauzner-Bechcicki$^{2}$, J. Zakrzewski$^{4,5}$, M. Lewenstein$^{3,6}$, F. He$^{1,7}$, J. Biegert$^{3,6}$ and M. F. Ciappina$^{8,9,10}$}

\address{$^1$Key Laboratory for Laser Plasmas (Ministry of Education) and School of Physics and Astronomy,
Collaborative innovation center for IFSA (CICIFSA), Shanghai Jiao Tong University, Shanghai 200240, China}
\address{$^2$Instytut Fizyki imienia Mariana Smoluchowskiego, Uniwersytet Jagielloński, Łojasiewicza 11, 30-348
Kraków, Poland}
\address{$^{3}$ICFO - Institut de Ciencies Fotoniques, The Barcelona Institute of Science and Technology, Av. Carl
Friedrich Gauss 3, 08860 Castelldefels (Barcelona), Spain}
\address{$^4$Institute of Theoretical Physics, Jagiellonian University in Krakow, Łojasiewicza 11, 30-348 Kraków,
Poland}
\address{$^5$Mark Kac Complex Systems Research Center, Jagiellonian University, Łojasiewicza 11, 30-348 Kraków,
Poland}
\address{$^{6}$ICREA, Pg. Lluís Companys 23, 08010 Barcelona, Spain}
\address{$^{7}$CAS Center for Excellence in Ultra-intense Laser Science, Shanghai 201800, China}
\address{$^{8}$Physics Program, Guangdong Technion - Israel Institute
of Technology, Shantou, Guangdong 515063, China}
\address{$^{9}$Technion -- Israel Institute of Technology, Haifa, 32000, Israel}
\address{$^{10}$Guangdong Provincial Key Laboratory of Materials and Technologies for Energy Conversion, Guangdong Technion – Israel Institute of Technology, Shantou, Guangdong 515063, China}
\ead{marcelo.ciappina@gtiit.edu.cn}

\vspace{10pt}
\begin{indented}
\item[]\today
\end{indented}

\begin{abstract}
We present a joint experimental and theoretical study of non-sequential double ionization (NSDI) in argon driven by a 3100-nm laser source. 
The correlated photoelectron momentum distribution (PMD) shows a strong dependence on the pulse duration, and the evolution of the PMD can be explained by an envelope-induced intensity effect. Determined by the time difference between tunneling and rescattering, the laser vector potential at the ionization time of the bound electron will be influenced by the pulse duration, leading to different drift momenta.
Such a mechanism is extracted through a classical trajectory Monte Carlo-based model and it can be further confirmed by quantum mechanical simulations.
This work sheds light on the importance of the pulse duration in NSDI and improves our understanding of the strong field tunnel-recollision dynamics under mid-IR laser fields.
\end{abstract}

\submitto{\NJP}

\section{Introduction}
The rapid and steady development of strong laser technologies has opened the door to studying ultrafast electron dynamics \cite{Rev2009,Rev2012,Amini_2019}. Ubiquitous in many strong-field phenomena, electron correlations during atomical multi-electron ionization play an instrumental role and have been widely investigated. Even the direct single-photon double ionization, in which the electron-electron interaction happens either before (shake off) or after (knock out) the photon absorption takes place, carries rich physical information~\cite{DPI2013,DPI2021}. If the electron emitted after single ionization is driven back to the nucleus by the intense laser, the inelastic scattering dynamics will be induced by the recollision, and nonsequential double ionization (NSDI) may occur. Early experiments measuring intensity-dependent ion yields show a famous knee structure \cite{knee1983,knee1994,knee1995}, emphasizing the importance of electron correlations during strong-field double ionization. Considering the recollision process, the bound electron can be knocked out immediately if the rescattering energy is larger than its ionization potential (electron impact ionization). Inversely, the bound electron may be excited and released in a subsequent laser field cycle. This process is called recollision excitation with subsequent ionization (RESI)~\cite{corkum1993,kulander1995,RESI2001}. Hence, the ionization time difference of the two electrons can help us understand the distinct NSDI mechanisms. To extract the time difference and better study the electron correlations, we measure the correlated photoelectron momentum distribution (PMD) with a Reaction Microscope (ReMi) based on the cold-target recoil-ion momentum spectroscopy \cite{MOSHAMMER1996,DORNER2000}. The ionization time can thus be mapped into the drift momentum under the strong laser, while the PMD can reveal the correlated drift momentum distribution. Previous studies of the finger-like patterns \cite{Rudenko2007,Ye2008}, the anti-correlated PMD \cite{antiPMD2008,antiPMD2011zhou,antiPMD2011Bondar,antiPMD2018} and other specific PMD structures \cite{Bergues2012,Sun2014}, have greatly improved our understanding of NSDI. Thus, one may expect that PMD could still shed light about unexplored NSDI dynamics.

Although NSDI has been widely studied under laser fields with a wavelength of around 800 nm \cite{Rudenko2007,Ye2008,Staudte2007,Chen2010,Ye2015}, there exist few studies under mid-IR laser fields~\cite{Wolter2015}. When we increase the laser wavelength, the ponderomotive energy ($U_p=E_0^2/4\omega^2$, where $E_0$ and $\omega$ are the electric field peak amplitude and the laser central frequency, respectively) of the laser field becomes larger, leading to the suppression of the RESI mechanism as the rescattering energy (around 3.17$U_P$) becomes larger \cite{Alnaser2008,Herrwerth2008}, and thus the two electrons tend to be emitted directly after the high-energy recollision, leading to the same drift momentum in the PMD. A near-axis V-shaped structure in the PMD under a 3100-nm laser field was shown in Ref.~\cite{Wolter2015}. The mechanism behind such structure can be extracted through a classical trajectory Monte Carlo model (CTMC), in which the multi-return rescattering and the asymmetric energy distribution are important \cite{Huang2016}. The investigation of NSDI under mid-IR laser fields can help us to better disentangle both different ionization mechanisms \cite{shaaran2019} and strong-field non-dipole effects~\cite{chen2020nondipole}. Upon now, we have understood the main mechanisms of NSDI under mid-IR laser fields, however, how will the PMD evolve when the mid-IR laser field changes its parameters, such as the pulse duration, remains still unexplored. Studies based on the evolution of the PMD, just like NSDI in near 800 nm laser fields \cite{Bergues2012,evoPMD2012,evoPMD2016}, can advance our knowledge about the strong-field NSDI dynamics.

In this work, in contrast to the previous investigation~\cite{shaaran2019}, we focus on the dependence on the pulse-duration evolution of the PMDs in Ar driven by a 3100-nm laser field with a fixed laser intensity of about  $8.0 \times 10^{13} \mathrm{~W} / \mathrm{cm}^{2}$. Such a relatively low intensity can help us to suppress the RESI mechanism and make the direct (e,2e) NSDI the dominant path. Thus, contrary to previous works \cite{Dong2016,Liu2014,xu2015}, our pulse-duration dependent results do not rely on the RESI mechanism and the multi-recollision channel. With the help of a semi-classical model, we show that the PMD evolution results from an envelope-induced intensity effect. This mechanism can be further confirmed by quantum simulations. Both our classical and quantum simulations show an excellent agreement with the experimental results. With a complete comprehension of the underlying physics, we aim for a deeper understanding of the ultrafast tunnel-recollision dynamics in NSDI. Atomic units are used throughout, unless otherwise stated.

\section{Experimental setup}

The experiment is conducted with our 160--kHz repetition rate, 3100--nm optical parametric chirped pulse amplification (OPCPA) \cite{Chalus2009,Baudisch2015,Elu2021} light source whose pulses are sent into a high-resolution ReMi; see Ref.~\cite{Wolter2015} for details. The OPCPA allows us to reach intensities in the $10^{14}\,\rm{W/cm^2}$ range, thus placing the investigation into the quasistatic (QS) regime with Keldysh parameters $\gamma<0.3$~\cite{keldysh1965}. The high repetition rate of the OPCPA~\cite{Chalus2009,Thai2011} is imperative to achieve acceptable signal--to--noise by countering the unfavorable $\lambda^{-4}$ scaling of electron rescattering \cite{Colosimo2008}. The measurements are performed on a supersonic cold Argon gas target for three different laser pulse durations,``4T'', where T is the laser period (FWHM = 41 fs), ``6.5T'' (FWHM = 68 fs) and ``13.5T'' (FWHM = 140 fs). Key to the measurements is the ReMi's ability to record 3D particle momentum distributions. Ins short, combined static electric and magnetic fields guide electrons and ions in opposite directions onto multi-hit capable delay line detectors in which the arrival time is translated into position information~\cite{MOSHAMMER1996,DORNER2000}. Note that we have used this system in our previous study~\cite{Wolter2014}, at similar conditions with laser peak intensity of $9\times10^{13}$ W/cm$^2$, to investigate the very low energy structures (VLES) and discussed the involvement of Rydberg states with the recapturing of tunnel-ionized electrons, leading to the zero energy structure (ZES).

\section{Theoretical models}
\subsection{Semi-classical model}

The classical trajectory Monte Carlo 
model \cite{Leopold1979,ctmc1994} has been widely used to study double ionization (DI) of atoms in strong laser fields
\cite{Ye2008,HoCTMC2005,agapiCTMC2008}. The laser field in our classical model is expressed as $\mathbf{E}(t)=E_{0}f(t)\sin(\omega t) \,\hat{\mathbf{x}}$, where the pulse envelope $f(t)=\sin^{2}(\frac{t\pi}{\tau})$, $\tau$ is the pulse duration, $E_{0}$ is the laser peak amplitude, and $\omega$ is the laser central frequency. 
The Keldysh parameter \cite{keldysh1965} in this work is smaller than 1, and thus we can use the ADK theory \cite{ADK1991} to describe the tunneling step of the first laser-ionized electron.
The tunneling rate can then be expressed as 
\begin{equation}
W_{0}\left(t\right)=4\left(\frac{4 I_{p}^{(1)}}{E\left(t\right)}\right)^{\frac{2}{\sqrt{2I_{p}^{(1)}}}-1} \exp \left[\frac{-2 \left(2I_{p}^{(1)}\right)^{\frac{3}{2}}}{3 E\left(t\right)}\right],
\end{equation}
where the first ionization potential is fixed to $I_{p}^{(1)}=0.579$ a.u. 
The initial conditions are set as follows: the
position of the first tunneling electron is approximated by $\mathbf{x}_{0}=-\frac{I_{p}^{(1)}}{|E(t)|} \hat{\mathbf{x}}$, its
transverse momentum
is approximated by a Gaussian distribution with width $\sqrt{\frac{|E(t)|}{2 \sqrt{2 I_{p}^{(1)}}}}$ and the longitudinal momentum is assumed to be zero, while the initial condition for
the second bound electron is determined by a canonical ensemble \cite{canonical1982}, in which its energy equals the second ionization potential of Ar ($I_{p}^{(2)}=1.016$ a.u.). 
The sampling number at different tunneling times $t_i$ is assigned according to the weight $W_{0}(t_i)$.
Having set the initial ensemble, the classical equations of motion are then solved:
\begin{equation}
	\label{equ1}
	\frac{d \mathbf{r}_{i}}{d t}=\frac{\partial H}{\partial \mathbf{p}_{i}}, \, \frac{d \mathbf{p}_{i}}{d t}=-\frac{\partial H}{\partial \mathbf{r}_{i}}.
\end{equation} 
Here, the Hamiltonian of the two-electron system can be written as
\begin{equation}
	\label{equ2}
	H=\sum_{i=1}^{2}\left[\frac{1}{2} p_{i}^{2}-\frac{2}{\sqrt{r_{i}^{2}+s}}\right]+\frac{1}{\left|\mathbf{r}_{1}-\mathbf{r}_{2}\right|} +\sum_{i=1}^{2} \mathbf{r}_{i} \cdot \mathbf{E}(t),
\end{equation}
where the parameter $s$ is set to $0.1$ to avoid the Coulomb singularity at the origin. 
In our classical model, the bound electron in the initial ensemble is marked as $e_{1}$ and the tunneling electron is marked as $e_{2}$.
After $e_{2}$ tunnels out at $t_i$, we define the instant at which $e_{2}$ returns back and the two electrons are closest as the recollision time $t_{re}$. Subsequently, we define the instant at which the energy of the bound electron becomes larger than 0 and the electron is not trapped afterwards at the ionization time $t_{ion}$ of $e_1$.
In total, more than $10^{5}$ double ionization events are collected to guarantee the convergence of the results.




\subsection{Quantum models}

The quantum mechanical solution for two-electron systems interacting with a long-wavelength strong laser field is a formidable numerical problem, even when the electron spin is neglected.
To overcome this difficulty, reduced-dimensional models, which have long proved their usefulness in the single active electron (SAE) case~\cite{corkum1993,Eberly1989a,Mandrysz2019}, have made their way toward multi-electron studies~\cite{Ruiz2003,Ruiz05,Efimov21,Thiede18,Efimov18,Becker2012,Eckhardt2008}.
In the case of two-electron systems, the simplification involves a reduction from 3 spatial dimensions per electron (3+3) to 1 spatial dimension per electron (1+1).

\begin{figure}[t]
	\centering
	\includegraphics[width=0.9\textwidth]{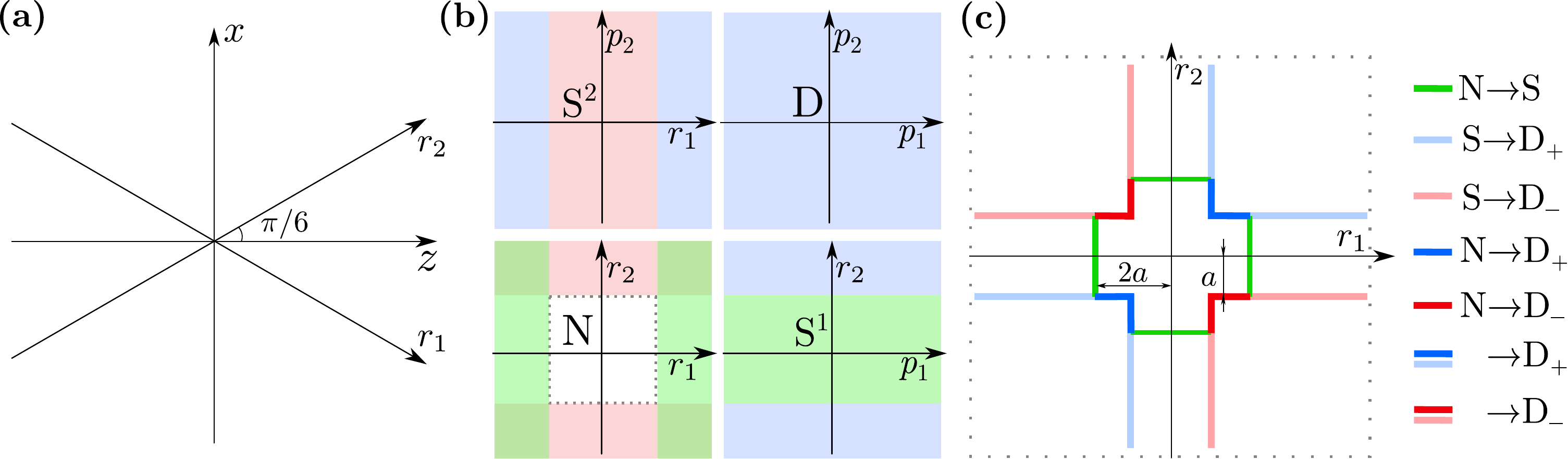}
	\caption{(a) Geometry of the ES model. The saddle tracks forming the directions ($r_1$, $r_2$) along which electrons are allowed to move. The polarization axis points along $z$. (b) Multiple grids (regions) used to hold the wavefunction of the system. Region $N$ approximating the neutral atom, regions $S^{1}$, $S^{2}$ approximating singly charged ions (ionized in the $r_1$, $r_2$ directions respectively) and a common region approximating doubly charged ions $D$ (for details see the main text). (c) Definitions of flux boundaries located in the dotted area of region $N$. The parameter $a$ is equal to the classical quiver amplitude $E_0/\omega^2$.} 
	\label{FigES}
\end{figure}

Apart from the technical advantages of such models, the dimensional reduction comes with additional freedom to select the axes (geometry).
The most straightforward choice of the reduced-dimensional axes in the case of linear polarization laser fields is the so-called Eberly or Rochester (EB) model~\cite{Su1991,Javanainen1988,Grobe92}. In this approach, each electron is left on one axis exactly parallel to the polarization axis. 
This choice makes it easier to compare with experimentally available information. However, the downside is that the e-e Coulomb interaction is overestimated. This problem can be remedied in the context of reduced-dimensional models with a different choice of the simulation axes.
One such choice is given by the classical Eckhardt-Sacha (ES) model~\cite{Eckhardt_2006}, where each electron travels on a one-dimensional track inclined with respect to the polarization axis at a constant angle. The choice of the inclination angle in this model is derived from the existence of a saddle in the electrons potential energy during simultaneous escape, and as such is better for modeling the (e,2e) NSDI ionization pathway (see Fig.~\ref{FigES}(a)).

The laser parameters in this work, i.e., moderate 
laser intensity, long wavelength, and long laser pulse duration, all result in a vast expansion of the wavefunction encoded in the position-space representation, which poses a difficulty for traditional grid-based methods.
The common method to deal with such problems is the t-SURFF technique, which has recently been successfully extended to two-electron systems \cite{zielinski2016double,armin2020}, although at shorter wavelengths. 
Here, we employ a different method, which involves coherent transfer of the (multi-)electron wavefunction~\cite{lein2000intense,prauzner2007time,prauzner2008quantum}.
In this approach, the electron wavepacket, once sufficiently far from the core, is coherently transferred into separate, non- or semi-interacting regions in which the evolution is performed in momentum or mixed representations, preventing its absorption and hence the loss of information.
In the following, we describe the details of the method.

The physical space is divided into four identically sized grids (regions) - the main region ($N$) hosting the initial wavefunction and the auxiliary, two single-ionized ($S^1$, $S^2$) and one double-ionized ($D$) regions (see Fig.~\ref{FigES}(b)).
The $S^1$, $S^2$, and $D$ regions store wavefunctions that evolve in parallel and similarly to the $N$ region using the split operator method, except that the evolution in the coordinate corresponding to the freed electron is performed solely in the momentum space without the Coulombic interaction (Volkov evolution). The free (ionized) electron direction is marked by the upper script in the $S$ region symbol. In the case of the $D$ region, the wavefunction evolves without any Coulombic interaction in the momentum space. 
The coherent wavepacket transfer process is cascading, i.e., occurs in two steps:
In the first step, parts of the wavefunction from the $N$ region, which occupy the space past half of the grid size in the $i$ direction ($i={1,2}$), are subtracted from it and transferred to the $S^i$ region, stored in mixed representations: $\psi(r_2,p_1)$ for $S^1$ and $\psi(p_1,r_2)$ for $S^2$ (light green and red parts of the $N$ region in Fig.~\ref{FigES}(b)).
In the second step, parts of the wavefunction from the $S^i$ region extending past half of the grid size in the $i$ direction are transferred to the $D$ (doubly-ionized) region stored in the $\psi(p_1,p_2)$ representation (light blue parts of the $S^1$ and $S^2$ regions in Fig.~\ref{FigES}(b)).

The description above can be summarized by Eqs.~(\ref{q-ham-1})-(\ref{q-ham-3}) in which the Hamiltonians $H_N$, $H_S^i$, $H_D$ correspond to the regions $N$, $S^i$, $D$, respectively:
\begin{eqnarray}
	\hat{H}_D &=& \sum_{i} \frac{1}{2} \hat{p}_{i}^{2} + A(t) \hat{p}_i \label{q-ham-1} \\
	\hat{H}_S^{i} &=&  \hat{H}_D  -\sum_{j\neq i} \frac{2}{\sqrt{\hat{r}_{j}^{2}+ \epsilon}} \label{q-ham-2} \\
	\hat{H}_N &=&  \hat{H}_D + H_{ee} -\sum_{i} \frac{2}{\sqrt{\hat{r}_{i}^{2}+ \epsilon}}. \label{q-ham-3}
\end{eqnarray}
The electron-electron interaction part $H_{ee}$ depends on the applied model described above. For the EB model, the interaction is described by $H_{ee}^{EB}$ (Eq.~(\ref{q-model-EB})), while for the ES model, it takes the form $H_{ee}^{ES}$ (Eq.~(\ref{q-model-ES})), i.e.
\begin{eqnarray}
H_{ee}^{EB} &=& \frac{1}{\sqrt{(\hat{r}_{1}-\hat{r}_{2})^2+\epsilon}} \label{q-model-EB}\\
H_{ee}^{ES} &=& \frac{1}{\sqrt{(\hat{r}_{1}-\hat{r}_{2})^2+\hat{r}_{1} \hat{r}_{2}+\epsilon}}. \label{q-model-ES}
\end{eqnarray}

In addition to the efficient multi-region evolution, we also collect information about the quantum current fluxes streaming through boundaries located in the $N$ region (i.e. before the coherent transfer takes place; see Fig.~\ref{FigES}(c)). Such boundaries are located at distances proportional to the classical quiver amplitude $E_0/\omega^2$ and inherit their naming from the previously defined grid regions, except for an addition of the $+/-$ sign signifying whether the flux was along the field axis (correlated) or perpendicular to it (anti-correlated) towards the $D$ region. 

\section{Results and discussion}
\subsection{Experimental results}

\begin{figure}[h!]
	\centering
	\includegraphics[width=0.95\textwidth]{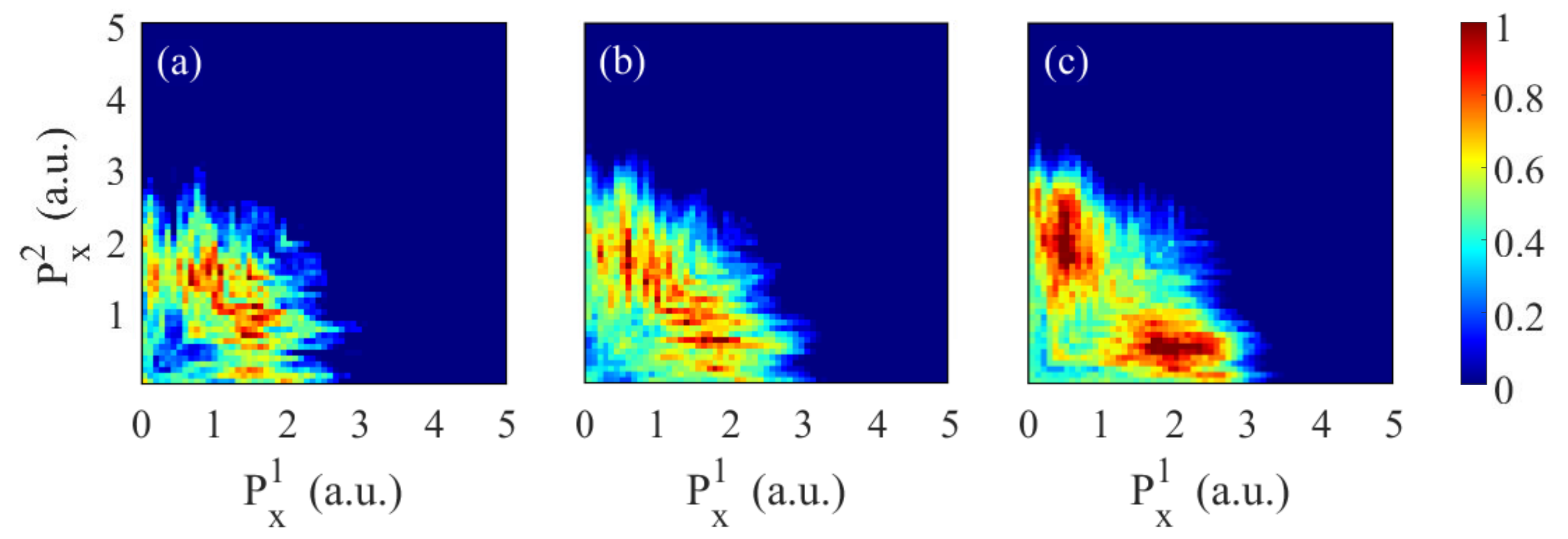}
	\caption{Experimental normalized correlated PMD along the polarization direction. The FWHM of the laser is 4T in (a), 6.5T in (b), and 13.5T in (c). The peak laser intensity is $9.0 \times 10^{13} \mathrm{~W} / \mathrm{cm}^{2}$ and the laser wavelength is 3100 nm.}
	\label{Fig2a}
\end{figure}

We record the 3D PMD taken in coincidence with the $\rm{Ar}^{2+}$ ions. Since the dead time particle detector is about 10 to 20 ns \cite{DORNER2000}, we detect only one electron in the first quadrant \cite{COLTRIMS2000,Ullrich_2003}. The other electron momentum can be reconstructed following the momentum sum conservation of the three involved particles coincidence, i.e.
\begin{equation}
    P_x^{Ar^{2+}}=-(P_x^1+P_x^2),  \;\;\;\;P_y^{Ar^{2+}}=-(P_y^1+P_y^2),  \;\;\;\;P_z^{Ar^{2+}}=-(P_z^1+P_z^2).
\end{equation}
By extracting the doubly charged ion longitudinal momentum $P_x^{Ar^{2+}}$ and one of the electrons $P_x^{1,2}$, we can deduce the other electron longitudinal momentum $P_x^{1,2}$. In order to ensure momentum conservation and realize a safe reconstruction of the missing particle, we filter out the number of coincidence events per laser pulse to be exactly one. In Figs.~\ref{Fig2a}(a)-(c), we show the ($e_1,e_2$) correlated PMD along the laser polarization axis and under different laser pulse durations. The PMD from the shortest pulse duration (see Fig.~\ref{Fig2a}(a)) shows correlated momenta with similar amplitude along the diagonal $P_x^1=P_x^2$. When increasing the pulse length, the two outgoing electron momenta start displaying an asymmetry as shown in the PMD from Fig.~\ref{Fig2a}(b). Finally, at longer pulse durations (see Fig.~\ref{Fig2a}(c)), the ($e_1,e_2$) correlated momentum map clearly exhibits a cross-like shape \cite{Wolter2015}. As a consequence, the experimental PMDs show a strong dependence on the pulse duration at a fixed laser intensity.

\subsection{Extracting mechanisms through the classical model}

\begin{figure}
	\centering
	\includegraphics[width=0.96\textwidth]{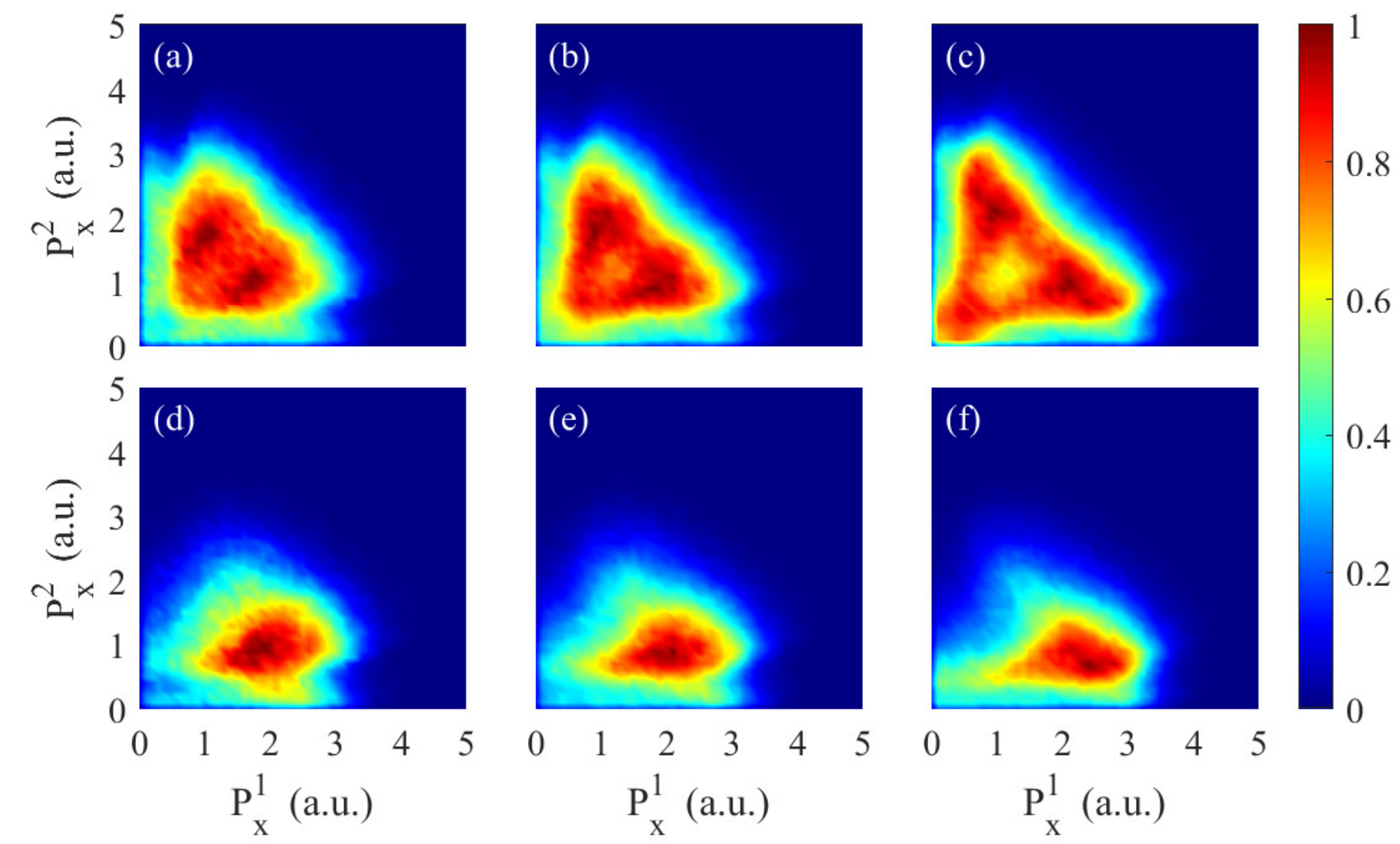}
	\caption{Normalized correlated PMD spectra along the polarization direction. In (a)-(c), the two electrons are indistinguishable. In (d)-(f), the two electrons are distinguishable, and $\mathrm{P}_{x}^{2}$ is the tunneling electron momentum. The pulse duration $\tau$ is 11T (FWHM=4T) in (a) and (d); 17T (FWHM=6T) in (b) and (e); 38T (FWHM=14T) in (c) and (f). The laser intensity is $8.0 \times 10^{13} \mathrm{~W} / \mathrm{cm}^{2}$ and the laser wavelength is 3100 nm.}
	\label{Fig1}
\end{figure}

The mechanisms of NSDI under long-wavelength laser pulses have already been studied with CTMC~\cite{Huang2016,chen2020nondipole}. The rescattering electron returns with a large momentum because of the high ponderomotive energy, however, emitting the bound electron just requires a small part of such rescattering energy. Thus, there is a clear asymmetry between the two electrons. With the help of CTMC, we can clearly understand the underlying dynamics. However, previous works \cite{Huang2016,shaaran2019} did not consider the influence of the pulse duration. In Figs.~\ref{Fig1}(a)-(c), we show that the PMDs have clear differences under different pulse durations and the results are consistent with the experimental results (see Fig.~\ref{Fig2a}). 

To understand the influence of the pulse duration on the NSDI mechanisms, we show the correlated PMD without symmetrization in Figs. \ref{Fig1}(d)-(f). In these PMD spectra, $e_{2}$ is the tunneling electron. The tunneling electron tends to have smaller momenta, and this can be understood by a classical approximation. The final momentum of the tunneling electron can be approximated by $p_{re}-A(t_{re})$, where $p_{re}$ and $A(t_{re})$ are the momentum and the vector potential at the recollision time, respectively. Right after the recollision, the tunneling electron still carries a large remaining momentum, which has an opposite direction compared to the vector potential, and thereafter it is slowed down by the laser field, leading to a small momentum in the final state. On the contrary, the bound electron acquires a small part of the rescattering energy, and the total energy of it after recollision is around zero. Thus, the bound electron is ionized with a small $p_{0}$ and the final momentum can be approximated by $-A(t_{0})$, where $p_{0}$ and $A(t_{0})$ are the momentum and the vector potential at the ionization time of $e_1$, respectively. For a high rescattering energy, the direct (e,2e) channel dominates, thus, $t_{re}$ and $t_{0}$ are close to each other and are around at the maximum of the vector potential. Hence, the final momentum of $e_1$ ($-A(t_{0})$) is larger than $e_2$ ($p_{re}-A(t_{re})$). The differences in Figs.~\ref{Fig1}(a)-(c) can be summarized by pointing out two main factors: the momentum of $e_{1}$ becomes larger as the pulse duration increases, while the momentum distributions of $e_{2}$ (the tunneling electron) only have slight changes. We will analyze in detail why the pulse duration has such an influence on the PMD of the two electrons next.

\begin{figure}
	\centering
	\includegraphics[width=0.6\textwidth]{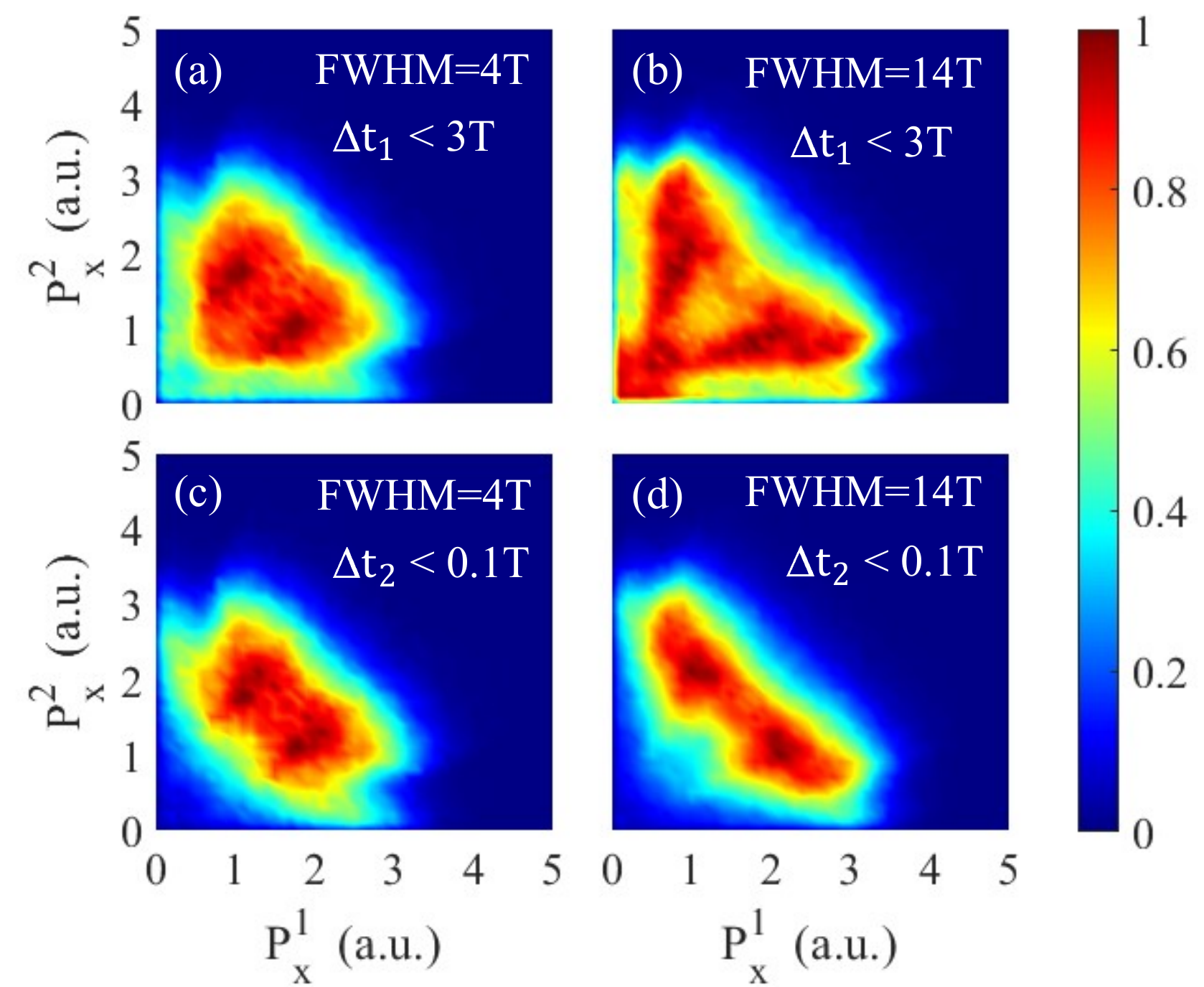}
	\caption{Normalized correlated PMD spectra. $\Delta t_{1}$ is the time difference between tunneling and recollision. $\Delta t_{2}$ is the time difference between recollision and the emitting of the bound electron.  (a)-(b) $\Delta t_{1}<3T$, (c)-(d) $\Delta t_{2}<0.1T$. The laser FWHM is 4T in (a) and (c) and 14T in (b) and (d).}
	\label{Fig2}
\end{figure}

The RESI mechanism \cite{Dong2016,jesus2004atomic} is always important when considering NSDI. Also, multi-return dynamics, in which the electron returns without a recollision due to the transverse displacement and the recollision may happen after several return times, is relevant when considering NSDI under long-wavelength laser pulses \cite{Huang2016}. Therefore, it is necessary to examine whether the two mechanisms are the reasons behind the experimental PMD spectra. For a longer-duration laser field, the tunnel-recollision process may contain more return times than a shorter pulse. To exclude the influence of these events, in which the number of returns is larger than the maximum value under the shorter pulse, in Figs.~\ref{Fig2}(a) and (b) we select the DI events in which the time difference between tunneling and recollision ($\Delta t_{1}$ is directly related to the multi-return trajectories \cite{Huang2016}) is smaller than 3T. However, we still can see a clear difference in the PMDs under different pulse durations. Thus, those multi-return events, in which the number of returns is larger in the longer-duration laser field, do not account for the experimental results. In Figs.~\ref{Fig2}(c) and (d), we select DI events in which the time difference between recollision and emission of the bound electron is smaller than $0.1T$ (excluding the influence of RESI). Here, the correlated PMD spectra still show a clear difference. Therefore, there is no direct relation between the experimental results and the RESI mechanism.

\begin{figure}
	\centering
	\includegraphics[width=0.6\textwidth]{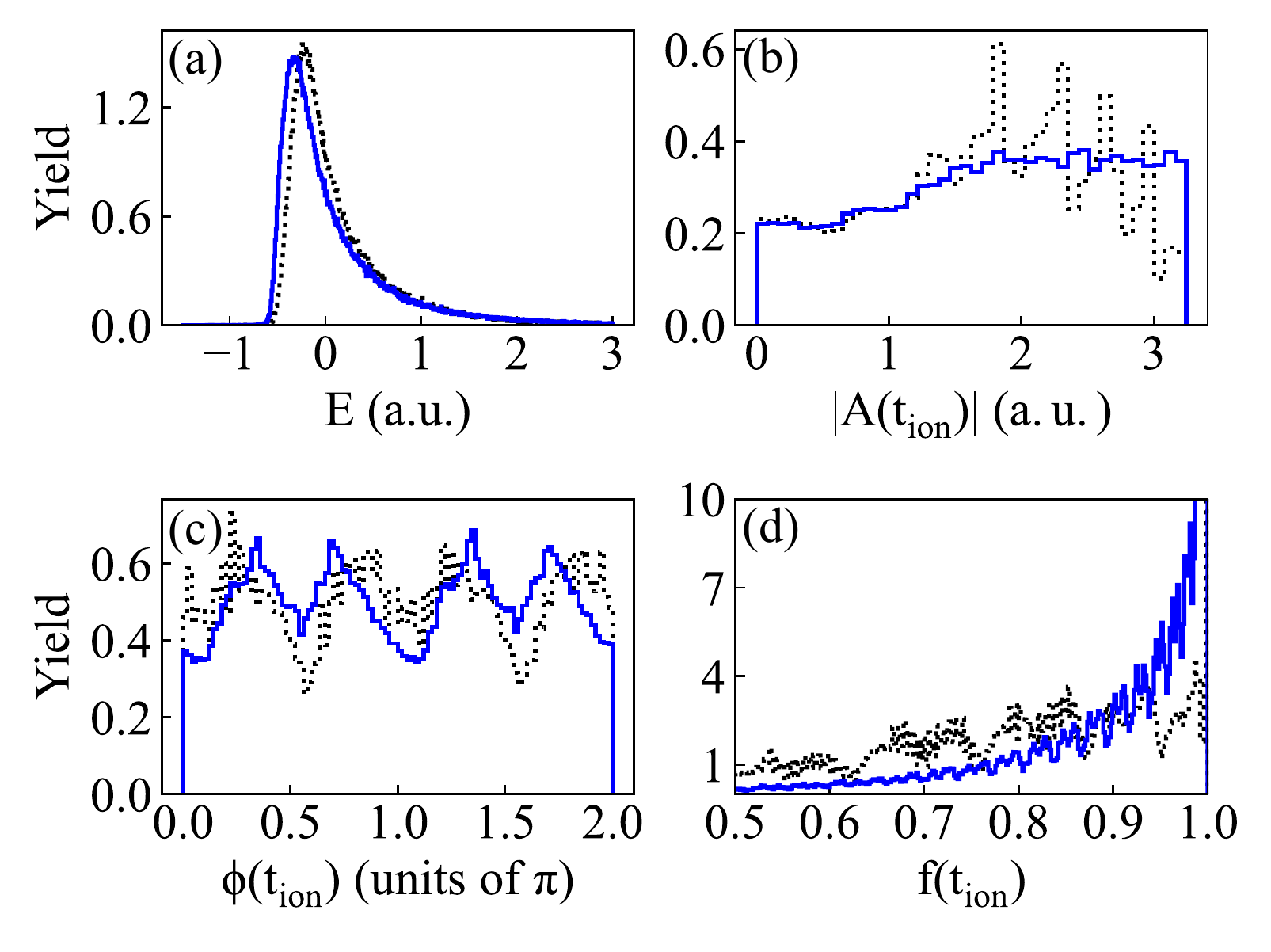}
	\caption{(a) The energy distribution of $e_{1}$ (the bound electron in the initial ensemble) right after the recollision (2 a.u. later). (b) The distribution of the absolute value of the vector potential at the ionization time of $e_{1}$ ($t_{ion}$). (c) The laser phase distribution at the ionization time. (d) The distribution of the value of $f(t_{ion})$, where $f(t)$ is the envelope function ($f(t)=\sin^{2}(\frac{t\pi}{\tau})$). The laser FWHM is 4T for the dotted black line and 14T for the solid blue line.}
	\label{Fig3}
\end{figure}

In Fig.~\ref{Fig3}(a), we show the energy distribution of $e_{1}$ right after the rescattering of $e_{2}$ for two different pulse durations. Both distributions are located around 0 a.u., and therefore we conclude that the initial momentum of $e_{1}$ is small at the ionization time $t_{ion}$, and the final momentum of $e_{1}$ can be approximated by $-A(t_{ion})$ in both cases. In Fig.~\ref{Fig3}(b), when the laser FWHM changes from 4T to 14T, the distribution of $|A(t_{ion})|$ changes from 2 a.u. to 2.5 a.u. (the most probable value).  This change can help us to directly understand why the momenta of $e_1$ become larger for longer pulse durations in Figs.~\ref{Fig1}(d)-(f). Then we just need to consider why $A(t_{ion})$ tends to become larger as the pulse duration increases. There are two possible factors: the phase of the vector potential at $t_{ion}$, or the envelope function $f(t_{ion})=\sin^{2}(\frac{t_{ion}\pi}{\tau})$ (e.g. the instantaneous laser intensity). In Fig. \ref{Fig3}(c), we show the phase distribution of $\omega\,t_{ion}$. If the phase distribution accounts for the dynamics behind Fig. \ref{Fig3}(b), the phase should be closer to 0 or $\pi$ to obtain a larger instantaneous vector potential when we increase the pulse duration. However, such a tendency is not observed in Fig. \ref{Fig3}(c), thus, we can not claim the phase distribution as the reason behind the PMD of Fig.~\ref{Fig3}(b). In Fig. \ref{Fig3}(d), the distributions of the value of $f(t_{ion})$ show a large difference under different laser pulse durations. When the laser FWHM is 14T, the distribution of $f(t_{ion})$ is located around 1.0. However, when the pulse duration is shorter, the distribution of $f(t_{ion})$ lies in the interval [0.7,1.0]. Thus, for the case of a shorter pulse, the instantaneous vector potential tends to be smaller than that of a longer pulse because of the envelope-induced intensity difference. Therefore, we can understand the PMD of $e_{1}$ under different pulse durations through Fig.~\ref{Fig3}(d).

\begin{figure}
	\centering
	\includegraphics[width=0.7\textwidth]{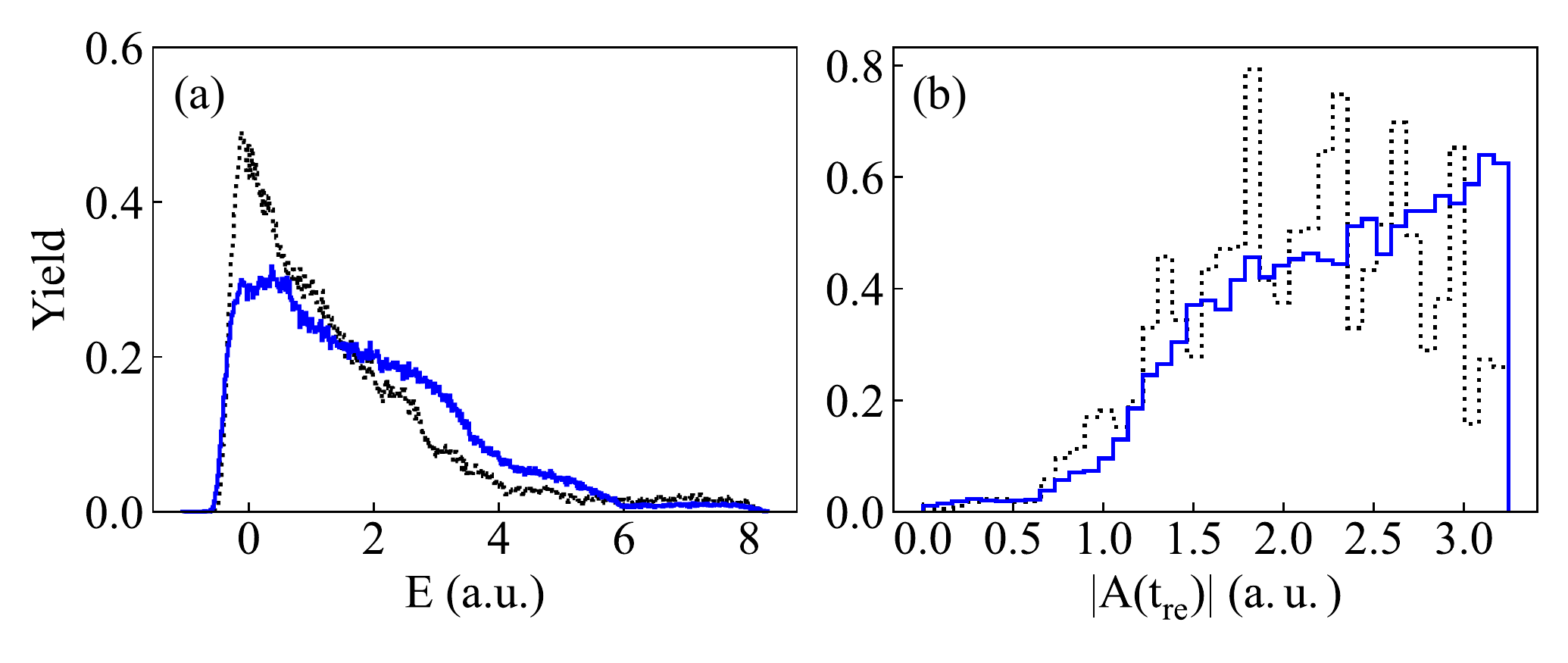}
	\caption{(a) The energy distribution of $e_{2}$ (the tunneling electron) right after the recollision (2 a.u. later). (b) The distribution of the absolute value of the vector potential at the recollision time of $e_{2}$ ($t_{re}$). The laser FWHM is 4T for the dotted black line and 14T for the solid blue line.}
	\label{Fig4}
\end{figure}

\begin{center}
\begin{table}
	\centering
	\caption{\label{table1} $\left<E\right>$ is the average energy of the tunneling electron right after the recollision. $|\left<A\right>|$ is the average absolute value of vector potential right after the recollision. The laser intensity is $8.0 \times 10^{13} \mathrm{~W} / \mathrm{cm}^{2}$ and the laser wavelength is 3100 nm.}
	\footnotesize
	\begin{tabular}{@{}cccc}
		\br
		$\tau$ (T) & $\left<E\right>$ (a.u.)  & $|\left<A\right>|$ (a.u.) & $\sqrt{2\left<E\right>}-|\left<A\right>|$ (a.u.) \\
		\mr
		11 & 1.42 & 2.09 & -0.41 \\
		17 & 1.50 & 2.14 & -0.40 \\
		38 & 1.77 & 2.26 & -0.38 \\
		\br
	\end{tabular}\\
\end{table}
\end{center}
\normalsize

When it comes to the tunneling electron $e_{2}$, the momentum distributions only have slight changes under different pulse durations. 
In Fig.~\ref{Fig4}(a), we can see that the tunneling electron tends to have a larger momentum right after the recollision for longer pulse durations. In Fig.~\ref{Fig4}(b), the distributions of the vector potential at the time of recollision $t_{re}$ resemble those in Fig.~\ref{Fig3}(b). Thus, the final momentum of $e_{2}$ can be approximated by $p_{re}-A(t_{re})$. Because both $p_{re}$ and $|A(t_{re})|$ tend to become larger as the pulse duration increases, the ultimate momentum of $e_{2}$ does not necessarily increase. To gain a better quantitative understanding, we calculate the expected energy of $e_{2}$ right after the recollision and show the results in Table~\ref{table1}. Here, $p_{re}$ is approximated by $\sqrt{2E}$ ($E$ is the energy of the tunneling electron $e_{2}$ right after the recollision). Estimates are consistent with the simulation results of Figs.~\ref{Fig1}(d)-(f). All results suggest that the instant laser field at rescattering, which is different due to the different values of the envelope function, distinctly affects the PMD spectra. Thus, the mechanism behind the evolution of the experimental results for different pulse durations can be attributed to an envelope-induced intensity effect.

\subsection{Quantum mechanical simulations}

\begin{figure}[t]
	\centering
	\includegraphics[width=1.0\textwidth]{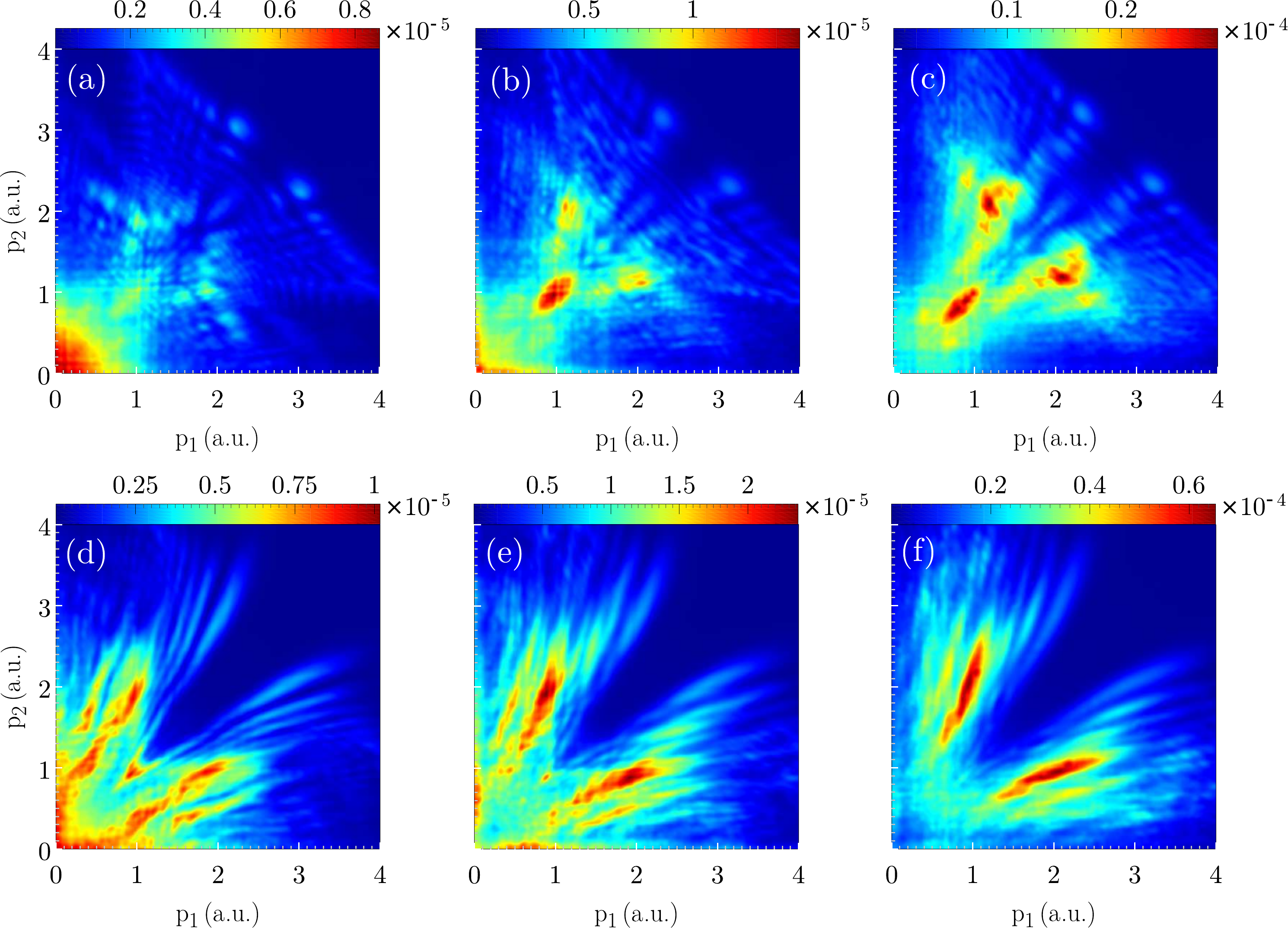}
	\caption{Results of the ES (top row) and EB (bottom) models for different FWHM Gaussian-enveloped pulse lengths: 4T (left), 6.6T (center), 13.5T (right). The plot coordinates correspond to momenta conjugate to the particular axes chosen by each model.}
	\label{wf2d_3100}
\end{figure}

The calculated PMDs of Argon exposed to 3100-nm pulses of variable length for the ES and EB models are shown in Fig.~\ref{wf2d_3100}.  
Evolving the wavefunction simultaneously on 4 separate grids, using the Hamiltonians from Eqs.~(\ref{q-ham-1})-(\ref{q-ham-3}) and allowing the system to relax through $0$-field post-propagation, accumulates the PMD information in the $D$ region presented in the plots. As stated before, in the case of the EB model, the $p$ axes are parallel to the polarization axis, while they are oblique in the ES model. Convergence near the $p$ axes is slow, and a cross-shaped gap is visible on all plots. In some of the previous works at shorter wavelengths (e.g., \cite{prauzner2008quantum}), an additional final coherent transfer (masked extraction and addition) steps from the $S$ and $N$ regions successfully completed the data without polluting the data with bound states.
For the wavelength considered in this work, applying such a technique with a fixed mask appears to introduce bound states for smaller FWHM pulse lengths (4T and 6.6T), while improving the dataset only slightly for both models at 13.5T FWHM pulse length (results not shown).
This feature is indicative of a substantial population of excited bound states at shorter pulse lengths. The general changes of the ES model observed in PMDs as the pulse length increases are consistent with the experimental data and the previously discussed classical results. For the EB model, the change in PMDs is less visible. The observed difference between the results obtained with the EB and ES models may be attributed to the fact that the ES model is better suited to modeling the (e,2e) NSDI ionization pathway. 

\begin{figure}[t]
	\centering
	\includegraphics[width=0.71\textwidth]{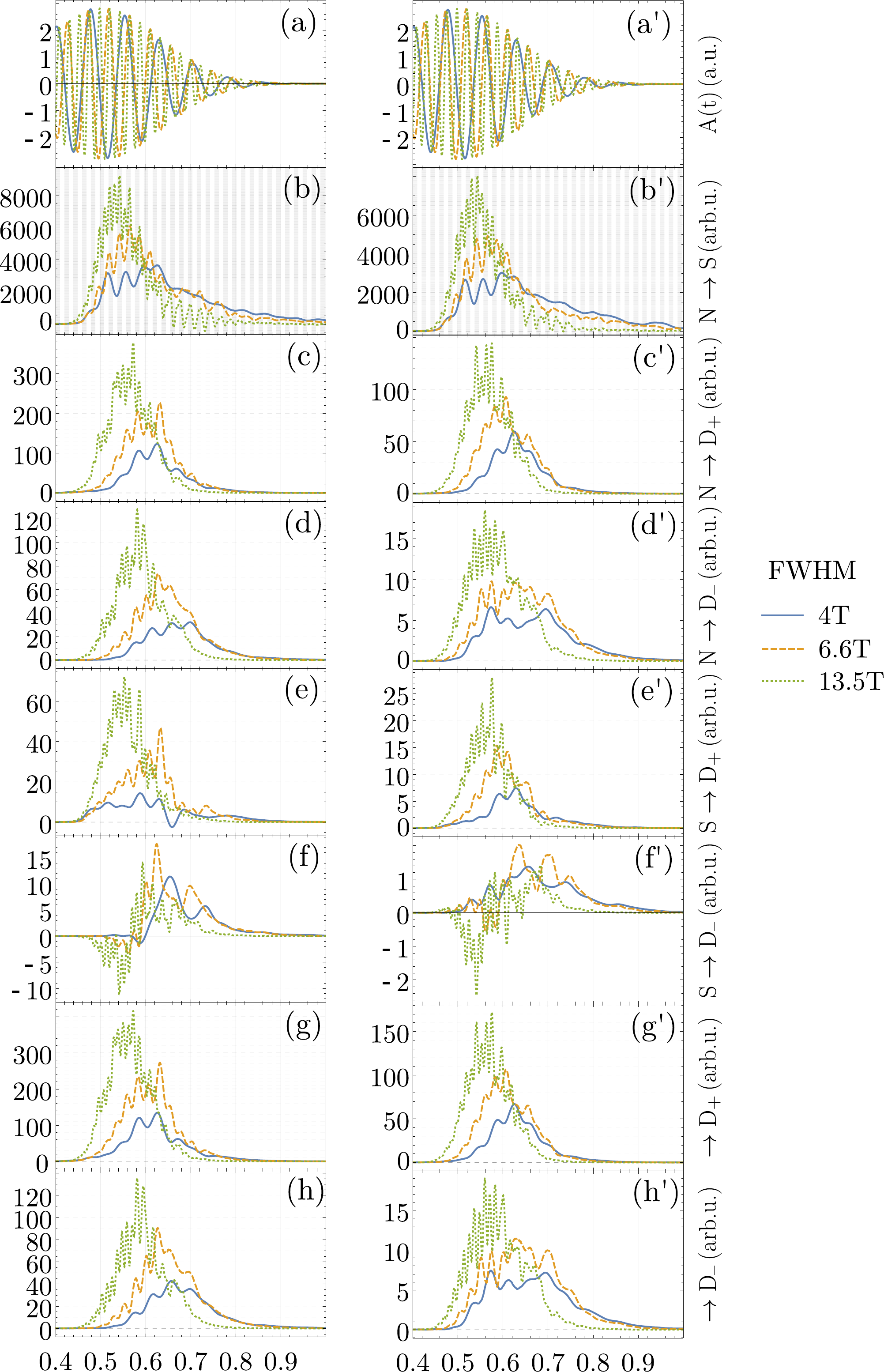}
	\caption{Vector potential ($a$) and probability fluxes at boundaries located at 2 times the quiver length for the ES (left column) and EB (right column, primed indexes) models. Flux from $N$ towards $S$ region ($b$), flux from $N$ towards $D$ region ($c$ - correlated/$d$ -anticorrelated), flux from $S$ towards $D$ region ($e$ - correlated/$f$ -anticorrelated) and cumulative flux towards $D$ region ($g$ - correlated/$h$ -anticorrelated). For clarity, the flux information was re-scaled per model.}
	\label{fluxes}
\end{figure}


The mechanism extracted in the classical case, confirming the PMD changes under different pulse durations due to different distributions of envelope values at the time of ionization (see Fig.~\ref{Fig4}(d)), also finds its counterpart in the quantum case. To show this effect, we shall analyze the probability fluxes (see Fig.~\ref{FigES}(c) for definitions and Fig.~\ref{fluxes} for results) that correspond to single ionization and various double ionization scenarios. The flux describing single ionization involves all possible events, that is, electrons that will return and recollide with the parent ion and electrons that will fly away to the detector. Regardless of the pulse length, single ionization peaks are located around the maximum of the pulse and throughout the pulse duration (and for some time after the pulse is gone). To infer the role of rescattering in DI, we calculated two fluxes populating the $D$ region, namely the one that directly crosses from the $N$ region and the other that comes from the 
$S$ region. Additionally, we are able to calculate fluxes to each of the PMD quadrants separately, i.e., to focus on both electrons escaping in the same direction (subscript $+$) or electrons escaping in opposite directions (subscript $-$). 
Looking at the envelope-aligned probability fluxes towards the $D$ region for different pulse lengths (see Figs.~\ref{fluxes}(c)-(h) and (c')-(h')) we notice a significant and consistent shift towards later times for shorter pulses. 
The shift of the maximum flux probability for the DI ($N\rightarrow D_{+}$) towards later times in case of shorter pulses confirms the envelope-induced intensity effect.

\subsection{Tunnel-recollision dynamics}

\begin{figure}
	\centering
	\includegraphics[width=1.0\textwidth]{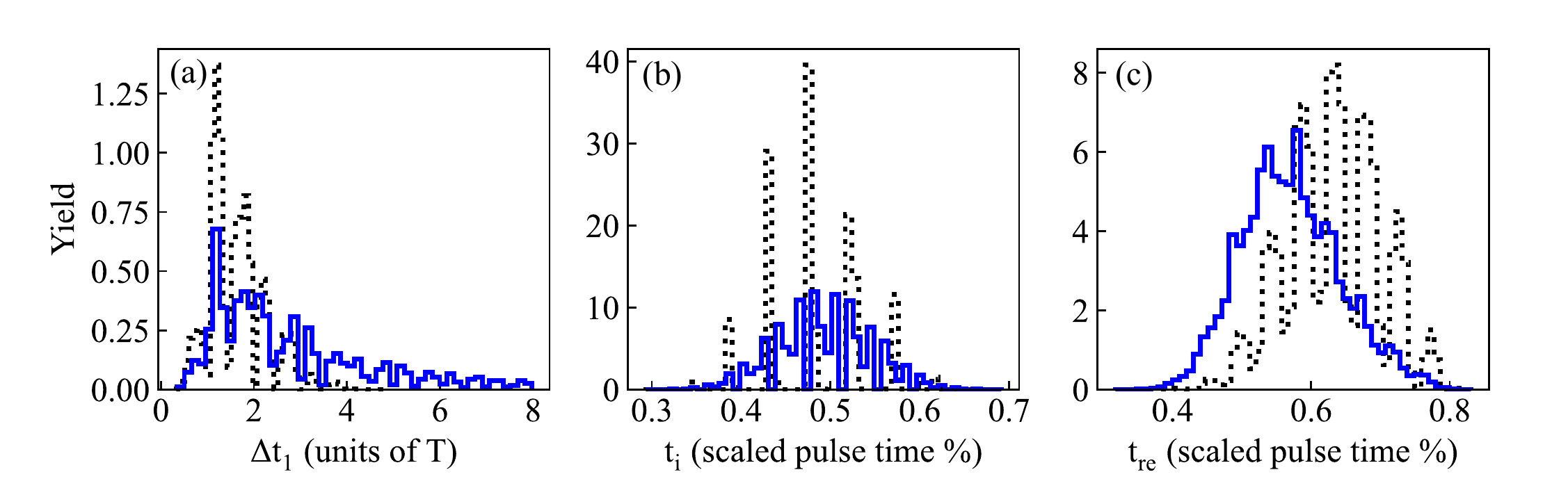}
	\caption{(a) The normalized distributions of the time difference between tunneling and recollision. (b) The normalized distributions of the tunneling time under the scaled pulse time (\%). (c) The normalized distributions of the recollision time under the scaled pulse time (\%). The laser FWHM is 4T for the dotted black line and 14T for the solid blue line. The data is collected through the CTMC model.}
	\label{Fig9}
\end{figure}

In essence, the envelope-induced intensity effect comes from the tunnel-recollision dynamics. In Fig.~\ref{Fig9}(a), we show the distribution of the time difference between tunneling and recollision. In both pulse durations, the time differences are most likely located around 1.2T, indicating that the multiple-return recollision process dominates in NSDI under 3100-nm laser fields~\cite{Huang2016} and the traveling time distributions are similar under different pulse durations. In Fig.~\ref{Fig9}(b), the tunneling time distributions under the scaled time are both located around the same value (center of the pulse). Considering the travel time after tunneling, the recollision most likely occurs 1.2T later. Thus, the recollision happens further back in the laser envelope with a smaller laser intensity for a shorter pulse in Fig.~\ref{Fig9}(c). Such an envelope-induced intensity effect under long-wavelength linear polarized lasers connects the PMD and the tunnel-recollision dynamics, a point that has not been discussed before.

\section{Conclusion and outlook}

In this work, we perform a joint theoretical and experimental study of NSDI comparing the PMDs in Ar under 3100-nm laser fields for varying pulse duration. Surprising unexplored duration-induced PMDs are found in which the signals approach the axes when increasing the pulse length. We demonstrate that there exists an envelope-induced intensity effect explaining the experimental results. Essentially, the magnitude of the vector potential at the time of recollision is affected by the pulse duration, leading to different drift momenta of the second laser-emitted electron. We further confirm the mechanism through quantum simulations. Both our classical and quantum results are consistent with the experimental data. Our findings indicate the importance of the pulse length effects for NSDI under quasi-static conditions. The found envelope-induced intensity effect may help to study tunnel-recollision dynamics through the information encoded in the PMDs.

\section*{Acknowledgements}
 This work was supported by Innovation Program of Shanghai Municipal Education Commission (2017-01-07-00-02-E00034), National Key R\&D Program of China (2018YFA0404802), National Natural Science Foundation of China (NSFC) (Grants No. 11925405 and No. 91850203). M.M., J. P.-B., J.Z. and M.L. acknowledge the National Science Centre, (Poland) Symfonia Grant No. 2016/20/W/ST4/00314.
 M. L. acknowledges support from ERC AdG NOQIA, Agencia Estatal de Investigación (“Severo Ochoa” Center of Excellence CEX2019-000910-S, Plan National FIDEUA PID2019-106901GB-I00/10.13039 / 501100011033, FPI), Fundació Privada Cellex, Fundació Mir-Puig, and from Generalitat de Catalunya (AGAUR Grant No. 2017 SGR 1341, CERCA program, QuantumCAT\_U16-011424 , co-funded by ERDF Operational Program of Catalonia 2014-2020), MINECO-EU QUANTERA MAQS (funded by State Research Agency (AEI) PCI2019-111828-2 / 10.13039/501100011033), EU Horizon 2020 FET-OPEN OPTOLogic (Grant No 899794), Marie Sklodowska-Curie grant STRETCH No 101029393. M. F. C. acknowledges financial support from the Guangdong Province Science and Technology Major Project (Future functional materials under extreme conditions - 2021B0301030005).
J.B. acknowledges financial support from the European Research Council for ERC Advanced Grant “TRANSFORMER” (788218), ERC Proof of Concept Grant “miniX” (840010), FET-OPEN “PETACom” (829153), FET-OPEN “OPTOlogic” (899794), FET-OPEN “TwistedNano” (101046424), Laserlab-Europe (871124), Marie Skłodowska-Curie ITN “smart-X” (860553), Symfonia, 2016/20/W/ST4/00314, MINECO for Plan Nacional PID2020–112664 GB-I00; AGAUR for 2017 SGR 1639, MINECO for “Severo Ochoa” (CEX2019-000910-S), Fundació Cellex Barcelona, the CERCA Programme/Generalitat de Catalunya, and the Alexander von Humboldt Foundation for the Friedrich Wilhelm Bessel Prize. J.B. and A.S. acknowledge additional support from Marie Sklodowska-Curie grant agreement No. 641272. Simulations were performed on the $\pi$ supercomputer at Shanghai Jiao Tong University.

\clearpage

\section*{References}
\bibliographystyle{iopart-num}
\providecommand{\newblock}{}


\end{document}